\documentclass[reprint, superscriptaddress, twocolumn, amsmath, amssymb, prx, longbibliography, floatfix]{revtex4-2}
\usepackage{graphicx}
\usepackage{dcolumn} 
\usepackage{bm}
\usepackage[colorlinks=true, citecolor=magenta, linkcolor=magenta]{hyperref}
\usepackage{mathrsfs}
\usepackage{color}
\usepackage{soul}
\usepackage{xcolor}

\definecolor{orange}{rgb}{1.0, 0.5, 0.0}
\definecolor{violet}{rgb}{0.78,0.08, 0.52}
\definecolor{green}{rgb}{0.11, 0.35, 0.02}
\definecolor{bluebell}{rgb}{0.64, 0.64, 0.82}
\definecolor{capri}{rgb}{0.0, 0.45, 0.73}

\usepackage{braket}

\usepackage{lipsum}
\usepackage{xargs}
\usepackage{setspace}

\begin{document}

\title{Digitizing ultrafast adiabatic passage with a pulse train}

\author{Bo Y. Chang}
\affiliation{School of Chemistry (CeT), Seoul National University, 08826 Seoul, Republic of Korea}
 
\author{Ignacio R. Sola}
\email{corresponding author: isolarei@ucm.es}
\affiliation{%
 Departamento de Qu\'imica F\'isica, Universidad Complutense (y Unidad Asociada I+D+i CSIC), 28040 Madrid, Spain
}%
\author{Svetlana A. Malinovskaya}%
\affiliation{%
Department of Physics, Stevens Institute of Technology, Hoboken, NJ 07030}

\author{Sebastian C. Carrasco}
\affiliation{DEVCOM Army Research Laboratory, 
Adelphi, MD 20783
}%

\author{Vladimir S. Malinovsky}
\affiliation{DEVCOM Army Research Laboratory, 
Adelphi, MD 20783
}%

\date{\today}

\begin{abstract} 
We present a digitized implementation of rapid adiabatic passage based on a train of weak, frequency-varying ultrafast pulses. Analytic conditions on the subpulse Rabi frequencies and detunings are derived to reproduce the continuous-time population dynamics of a conventional long-pulse excitation. We find that the reproduced dynamics achieves high fidelity even for pulse trains with a small number of subpulses, provided that each subpulse remains within the perturbative regime. The subpulses act as discrete samples of the underlying continuous evolution; consequently, more complex population dynamics, characterized by multiple oscillations prior to the onset of adiabaticity, require a larger number of subpulses for accurate reproduction. In addition, we demonstrate how the sidebands of a frequency comb can be exploited for resonant excitation at large carrier detuning and for the precise preparation of superposition states.
\end{abstract}


\maketitle

\section{Introduction}
Rapid Adiabatic Passage (RAP) is a well-known technique that allows for the robust preparation of Hamiltonian eigenstates or even, under more stringent conditions, superposition states~\cite{Brumer2003,SolaPRA1999,ChangPRA01,ChangJCP03,MalinovskyPRL2004,MalinovskyPRA04,MalinovskyPRA04b}. Typically, RAP is implemented using frequency-modulated ({\it i.e.,} chirped) laser pulses, although similar effects can be achieved in two-photon processes by appropriate pulse sequences as in STIRAP~\cite{GaubatzCPL1988,GaubatzJCP1990,OregPRA84,BergmannRMP98,ShoreCambridge2011}.
In both cases, measuring the interaction energy as the time-integrated Rabi frequency $\Omega(t)$, the robustness of the processes, incarnated in adiabatic conditions, demands larger pulse areas than the typical Rabi flopping so often used (because of its economy) in quantum state preparation. If Rabi flopping requires a $\pi$ area (an $h/2$ action), depending on the pulse features and the chosen adiabatic scheme, adiabaticity typically starts from $3\pi$.

Indeed, because adiabaticity is a threshold condition, the strength of the interaction can be as large as needed, depending on the Hamiltonian to which adiabatic passage is applied. For instance, one can distort the molecular potentials to assist in population transfer between states with very different geometries, located in different electronic states~\cite{GarrawayPRL98,SolaPRA00,SolaPRL00,MalinovskyJPCA03,JGVJPCA06,ChangJCP09,ChangJCP09b}. In the opposite limit, adiabatic passage has been used as an efficient ingredient in quantum information protocols~\cite{MalinovskyPRL2004,MalinovskyPRA04,MalinovskyPRA04,SolaAAMO18}.

In many cases, strong field interactions lead to unwanted effects in the quantum system, like alternative multiphoton processes that often outcompete the desired preparation process, or optical damage of the sample. It is therefore very convenient to find alternative robust procedures that reduce these adverse effects. With this in mind,
we have recently proposed the use of a train of pulses as an efficient implementation of pulse sequences that share some effects of adiabatic passage~\cite{SolaJPB22}.

A frequency comb is a powerful spectroscopic tool built by a succession of precisely-timed ultrafast pulses from mode-locked laser systems, creating a train of pulses~\cite{Ye05}. Frequency combs have a broad range of applications, among them precision optical metrology \cite{Hall2000,Telle99,Reichert99,StowePRL06,LiangPRA24}, atomic clocks \cite{Rosenband08} and coherent control of molecular dynamics~\cite{Shi_PRA10,Barmes_NatPhot13,HortonJOSAB13,LiuCPL16,Rubio_PRA18,FedoseevPRL2021,Ivanov_PRL22,Zhang_Photonics25}, including adiabatic passage in two-photon transitions while minimizing the population of the intermediate state, as in STIRAP~\cite{ShapiroPRA07,ShapiroPRL07,ShapiroPRL08,RangelovPRA12,Zheng_npj22,SolaJPB22,Yang2025_improving_STIRAP}.

The very different proposals in the literature using trains to simulate STIRAP have stirred some controversy regarding the precise mechanisms by which the trains operate.
In a recent contribution~\cite{SolaJPB22}, we have clarified the properties of the dynamics, showing analytically how to choose the train parameters to force the dynamics to emulate the behavior of a system under STIRAP evolution. We showed that the recipe can be extended to other STIRAP techniques, such as the generalized straddling STIRAP (S-STIRAP) scheme~\cite{MalinovskyPRA97} of population transfer in N-level system with sequential couplings.

In this work, we extend these ideas to create a digitized version of RAP using a single train of pulses, showing how individual pulses should be chosen such that its dynamics replicates those of the ``continuous'' evolution. We solve one of the main challenges of digitizing RAP: how to change the central frequency of the train sub-pulses. Besides increasing flexibility to avoid unwanted strong-field effects, digitizing RAP can also bring additional new possibilities. To illustrate this, we show how one can leverage the natural appearance of sidebands to create superposition states in multilevel systems, as explained in the paper's final section.

We envision that our work opens new avenues for quantum control. For example, Floquet engineering could be used to design structured subpulses that not only reproduce the target RAP dynamics, but also suppress unwanted sensitivities and dynamically decouple the system from extraneous degrees of freedom.

\section{Mimicking adiabatic passage dynamics with pulse trains: Analytic conditions}

\begin{figure}[t]
\includegraphics[width=0.8\linewidth]{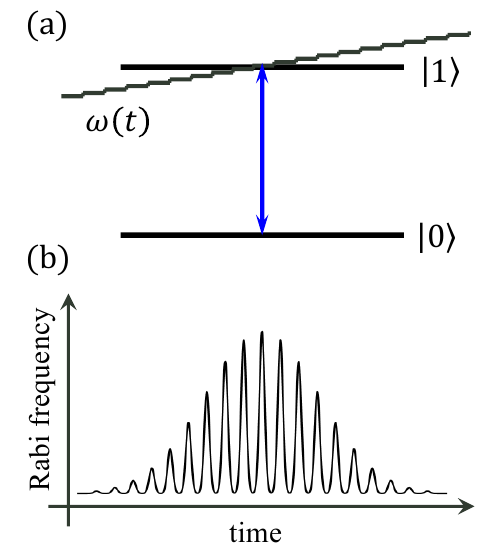}
\caption{(a) Energy level diagram of a two-level system. (b) Time-dependent Rabi frequency of the pulse train. For simplicity in the drawing, we show the case with $T=0$, whereas in most cases $T$ is much larger than the duration of each subpulse, $\tau$.}
\label{fig:1}
\end{figure}

In this section, we consider the interaction of a two-level (TL) system with a train of weak pulses (see Fig.~\ref{fig:1}) to reproduce the population dynamics of RAP, typically implemented using frequency chirping of a shorter transform-limited pulse~\cite{MalinovskyEPJD01,MalinovskyPRA01}, that is stretched in time. Our goal is to show how to choose pulse parameters for a train of short subpulses so that the system dynamics mimic those of a TL system driven by a long, henceforth called ``continuous'', pulse.

The general Hamiltonian for a two-level system within the rotating wave approximation (RWA) is given by
\begin{equation}
\tilde{H}(\tilde{t}) = -\frac{1}{2}\tilde{\Omega}(\tilde{t}) \sigma_1 
-\frac{1}{2}\tilde{\delta}(\tilde{t}) \sigma_3 , 
\end{equation}
where $\sigma_k$ denote the Pauli matrices, $\tilde{\Omega}(\tilde{t})$ is the Rabi frequency, and $\tilde{\delta}(\tilde{t})= \omega_{10} - \tilde{\omega}(t)$ is the detuning, which can be time-dependent ({\it e.g.} due to chirp modulation), $\omega_{10}$ is the transition frequency, and  $\tilde{\omega}(t)$ refers to the time-dependent field frequency. We will use tildes to refer to parameters of the continuous pulse, hence $\tilde{t}$ refers to the time where the ``continuous'' pulse is switched on, starting from $\tilde{t} = 0$, and $\tilde{\tau}$ is the pulse duration. Therefore, $\Delta\tilde{\delta} \approx
\tilde{\delta}(\tilde{\tau}) - \tilde{\delta}(0)$ is a measure of the pulse bandwidth or spectral range covered by the linearly chirped pulse, with time-dependent frequency defined as $\tilde{\delta}(t) = \tilde{\alpha} \left( \tilde{t} - \tilde{\tau}/2 \right)$, where $\tilde{\alpha}$ is the chirp rate. If necessary, the Hamiltonian can be conveniently extended to analyze the effect of
phase-controlled pulses, including complex phases, etc.

As a good approximation to the first order in time,
for a short interval of time $\Delta_t$ around $\tilde{t}_k$, 
such that both $\tilde{\delta}(\tilde{t})$ and $\tilde{\Omega}(\tilde{t})$ 
can be regarded as constant,
one can Taylor expand the time evolution operator
\begin{align}
\tilde{U}(\tilde{t}_k) &\approx \mathbb{I} - i \tilde{H}(\tilde{t}_k) \Delta_t \nonumber \\
&= \sigma_0 + \frac{i}{2}\tilde{\Omega}(\tilde{t}_k) 
\Delta_t \sigma_1 + \frac{i}{2}\tilde{\delta}(\tilde{t}_k) \Delta_t \sigma_3 \,.
\label{dUcont}
\end{align}
We would like to mimic the dynamic behavior of this system using a train of short subpulses of constant frequency, with Hamiltonian
$H = \sum_k h_k$ (piecewise sum over the $N$ subpulses of the train with a time period $T$) with
\begin{equation}
h_k(t) = -\frac{1}{2}\Omega_k(t) \sigma_1 -\frac{1}{2}\delta_k \sigma_3 , 
\end{equation}
where $\Omega_k(t)$ and ${\delta}_k$ are the Rabi frequency and (constant)
detuning of the $j$ subpulse of the train, peaking at time $t_k$. 
Figure~\ref{fig:1} shows a sketch of the system. We assume that the time separation between the subpulses, $T$, is much larger than the time duration of each subpulse, ${\tau}$. Variables without tilde will refer to parameters of the subpulses in the train.

The full-time evolution operator of the TL system interacting with the train can be decomposed into $N$ subunits (the number of subpulses in the train), each consisting of excitation by a subpulse, $\Pi_k$ followed by free evolution, ${u}_k$, such that ${U} = \Pi_k {u}_k$. We will use time-dependent first-order perturbation theory, where the evolution operators for free evolution and subpulse-driven evolution commute, so that whether to include free evolution before or after the subpulse action, or to split it in half between the two, will not affect the results. It is important, however, to consider in more detail the effect of free evolution, as it creates dynamical phases that may lead to changes not only in the phases but also in the populations of states during the dynamics.

For example, the time evolution operator in the Schr\"odinger picture is given by $U^S(t,t_0) = \Pi_k R_k(t) u_k R_k^{-1}(t_0)$, where $R_k(t)$ is a diagonal matrix with elements $|0\rangle\langle 0| + e^{-i\delta_k}|1\rangle\langle 1|$. When the detuning is constant, such that $\delta_k = \delta_{k+1} = \delta$, all $R_k$ matrices share the same phase factors. In this case, the $R_k$ and $R^{-1}_{k+1}$ elements cancel, and the full propagator acquires only an overall factor $e^{-i\delta (t-t_0)}$ acting on the $|1\rangle$ component. In contrast, for varying detuning, the transformations differ for each $R_k$, and this cancelation does not occur. Nevertheless, under the approximations valid for first-order perturbation theory, the free evolution contributes only phase factors and does not affect the populations.

Using the Magnus expansion to first order, we obtain
\begin{align}
{u}_k &\approx e^{-i\int_0^{T} {h}_k dt} = e^{i ({A}_k \sigma_1 
+ {\phi}_k \sigma_3)/2} \nonumber \\
&= \sigma_0 \cos ({A}_{ek}/2) 
+i \frac{{A}_k \sigma_1  + {\phi}_k \sigma_3}{{A}_{ek}} \sin ({A}_{ek}/2) \,,
\end{align}
where ${A}_k = \int_0^{T} {\Omega}_k(t)dt$ is the area of subpulse $k$,  ${\phi}_k =  \int_0^{T} {\delta}_k dt = T {\delta}_k$ is the optical phase acquired during a subunit of the train, and ${A}_{ek} = \sqrt{{A}_k^2 + {\phi}_k^2}$ is the effective pulse area.

For small areas of the subpulses and small detunings, we approximate the cosine/sine to first order in $A_{ek}$, which is equivalent to a first-order perturbative approximation of the dynamics, obtaining
\begin{equation}
{u}_k \approx \sigma_0 + \frac{i}{2} {A}_k \sigma_1 + \frac{i}{2} {\phi}_k \sigma_3 \ .
\label{utrain}
\end{equation}
Notice that $\phi_k = \delta_k T \ll 1$, where $T$ can be relatively large, which may more easily affect the validity of the approximation in non-resonant conditions. Including the free evolution with detuning, that is, the effect of the change in representation for different $u_k$, we evaluate 
\begin{align}
&R_k(t_{k+1}) u_k = R_k^\dagger(t_k) \nonumber \\
&=\left( \begin{array}{cc}
1 + \frac{i}{2} \phi_k & \frac{i}{2} e^{i\delta_k t_k} A_k \\ \frac{i}{2} e^{-i\delta_{k+1}t_{k+1}} A_k 
& e^{-i (\delta_{k+1}t_{k+1}-\delta_k t_k)} \left( 1 - \frac{i}{2}\phi_k\right) 
\end{array} \right)\,. \label{ujschro}
\end{align}
Now $\delta_{k+1}t_{k+1} = \delta_{k+1}t_k + \delta_{k+1}T = 
 \delta_{k+1}t_k + \phi_{k+1} \approx \delta_{k}t_k$, while
$\delta_{k+1}t_{k+1}-\delta_k t_k = (\delta_{k+1}-\delta_k) t_k + \phi_{k+1} \approx 0$.
The approximate values are obtained under the same approximations that lead to Eq.(\ref{utrain}), namely, $\phi_k \ll 1$. Hence, Eq.(\ref{ujschro}) gives
\begin{equation}
{u}_k \approx \sigma_0 + \frac{i}{2} \cos{\left(\delta_k t_k\right)} {A}_k \sigma_1 
- \frac{i}{2} \sin{\left( \delta_k t_k \right)} {A}_k \sigma_2 + \frac{i}{2}  {\phi}_k \sigma_3 \ ,
\label{utrainf}
\end{equation}
which is the same evolution operator in Eq.(\ref{utrain}) except for the addition of an extra phase. For instance, in the case
of full population inversion from $|0\rangle$ to $|1\rangle$, 
since $\sum_k^N \delta_k t_k = \tilde{\delta}\tilde{\tau}$,
the final state amplitude gets an overall phase of $e^{i (\pi/2 - \tilde{\delta}\tilde{\tau})}$
shifted from the usual $\pi/2$ phase. However, the phase does not affect the populations.

We want to find the parameters of the subpulses such that the population dynamics driven by the train mimics those of the continuous pulse. Because the train of pulses implies a constant spacing between the subpulses, in order to make Eq.(\ref{dUcont}) and Eq.(\ref{utrain}) equal, we need to divide the long continuous pulse into $N$ equi-separated slices sampled by the (equi-separated) $N$ subpulses of the train:
\begin{equation}
\mathrm{i)} \;\;\; {A}_k = \tilde{\Omega}(\tilde{t}_k)\Delta_t \hspace{0.5cm} \mathrm{and} 
\hspace{0.5cm} \mathrm{ii)} \;\;\; T {\delta}_k =  \tilde{\delta}(\tilde{t}_k) \Delta_t \ .
\label{matching}
\end{equation} 
If we write ${A}_k = {\Omega}_{k0} {\tau} S_0$, where ${\Omega}_{j0}$ is the peak Rabi frequency of subpulse $k$ and $S_0$ is the average value of the envelope of the subpulse (a shape factor) and consider the uniform sampling, $\Delta_t = \tilde{\tau} / (N-1)$, we find 
\begin{eqnarray}
{\Omega}_{k0} = \tilde{\Omega}(\tilde{t}_k) \frac{\tilde{\tau}}{S_0 {\tau} (N-1)} \label{matchgen1} \,, \\ 
{\delta}_k = \tilde{\delta}(\tilde{t}_k) \frac{\tilde{\tau}}{T (N-1)} \ .
\label{matchgen2}
\end{eqnarray}
To obtain the exact continuous time-evolution operator of Eq.~(\ref{dUcont}) including the phase displacement of Eq.(\ref{utrainf}), one would need to force another (demanding) condition, 
$\tilde{\delta}\tilde{\tau} = 2\pi n, n \in \mathbb{Z}$,
but this is typically not necessary, as the phases do not affect the populations.

For the simple case when the pulse duration of the ``continuous'' pulse is so long that one can actually make the whole pulse train to match that duration, then $\tilde{\tau} = (T+{\tau}) (N-1)$ and we can choose the subpulses to peak exactly at the sampled values of the continuous pulse,
such that $t_k = \tilde{t}_k$ and we can drop the tilde for the time variable. Then, calling $r_1 = T / {\tau}$ the ratio between the period of the train and the subpulse duration and $\Omega_{k0} = \Omega_k(t_k)$, we obtain
\begin{eqnarray}
{\Omega}_k(t_k) = \tilde{\Omega}(t_k) \frac{1 + r_1}{S_0} \approx \tilde{\Omega}(t_k) \frac{r_1}{S_0} \label{matchomega} \,, \\
{\delta}_k = \tilde{\delta}(t_k) \left( 1 + \frac{1}{r_1} \right) \approx   \tilde{\delta}(t_k) \ ,
\label{matchfreq}
\end{eqnarray}
where the approximations are valid when $r_1 \gg 1$, implicitly assumed when $T \gg \tau$. As shown by Eqs.(\ref{matchomega}) and (\ref{matchfreq}), to mimic the long-pulse dynamics using the pulse train, one just needs to sample the value of the frequency and amplitude of the pulse at $t_k$ ``discretizing'' the pulse with the subpulse, and simply scaling the amplitude of the subpulse taking into account the
amount of ``empty space'' in the train, which is not ``occupied'' by the subpulse, that is, the time-delay between the subpulses.

However, more typically, the duration of the train will be much longer than the duration of the chirped pulse, with several orders of magnitude of difference, {\it e.g.} picoseconds for the chirped pulse up to microseconds for the train. There are many possible implementations that scale the parameters such that Eqs.(\ref{matchgen1}) and (\ref{matchgen2}) are fulfilled. One may define the ratio between the duration of the long pulse and the accumulated duration of the subpulses in the train, $r_2 = \tilde{\tau} / N\tau$, such that
\begin{align}
\Omega_{k0} \equiv {\Omega}_{k}(t_k) &= \tilde{\Omega}(\tilde{t}_k) \frac{r_2}{S_0} \left(1 - \frac{1}{N-1} \right) \nonumber \\
&\approx \tilde{\Omega}(\tilde{t}_k) \frac{r_2}{S_0} \,, \label{otherlim1} \\ 
{\delta}_k &= \tilde{\delta}(\tilde{t}_k) \frac{r_2}{r_1} \left(1 - \frac{1}{N-1} \right) \nonumber \\
&\approx \tilde{\delta}(\tilde{t}_k) \frac{r_2}{r_1} \,.
\label{otherlim2}
\end{align}
The simplest solution is to make $r_2 = 1$, sampling the chirped pulse at $N$ equispaced times $\tilde{t}_k$, and scaling the time at which each subpulse is at its peak with the period of the train, such that $t_1 = \tilde{t}_1$ and $t_k = t_1 + T(k-1)$. More general solutions, with non-uniformed sampling of the continuous pulse by the frequency comb, are possible.

\section{Mimicking adiabatic passage dynamics with pulse trains: Numerical results}

We show some examples of how the dynamics of a train of pulses can be used
to discretize or digitalize the dynamics of a very long (``continuous'') pulse that serves as a guide, fixing the train parameters with
Eqs.(\ref{matchomega}) and (\ref{matchfreq}), with $\tilde{t} = t$.

\begin{figure}
\includegraphics[width=1.0\linewidth]{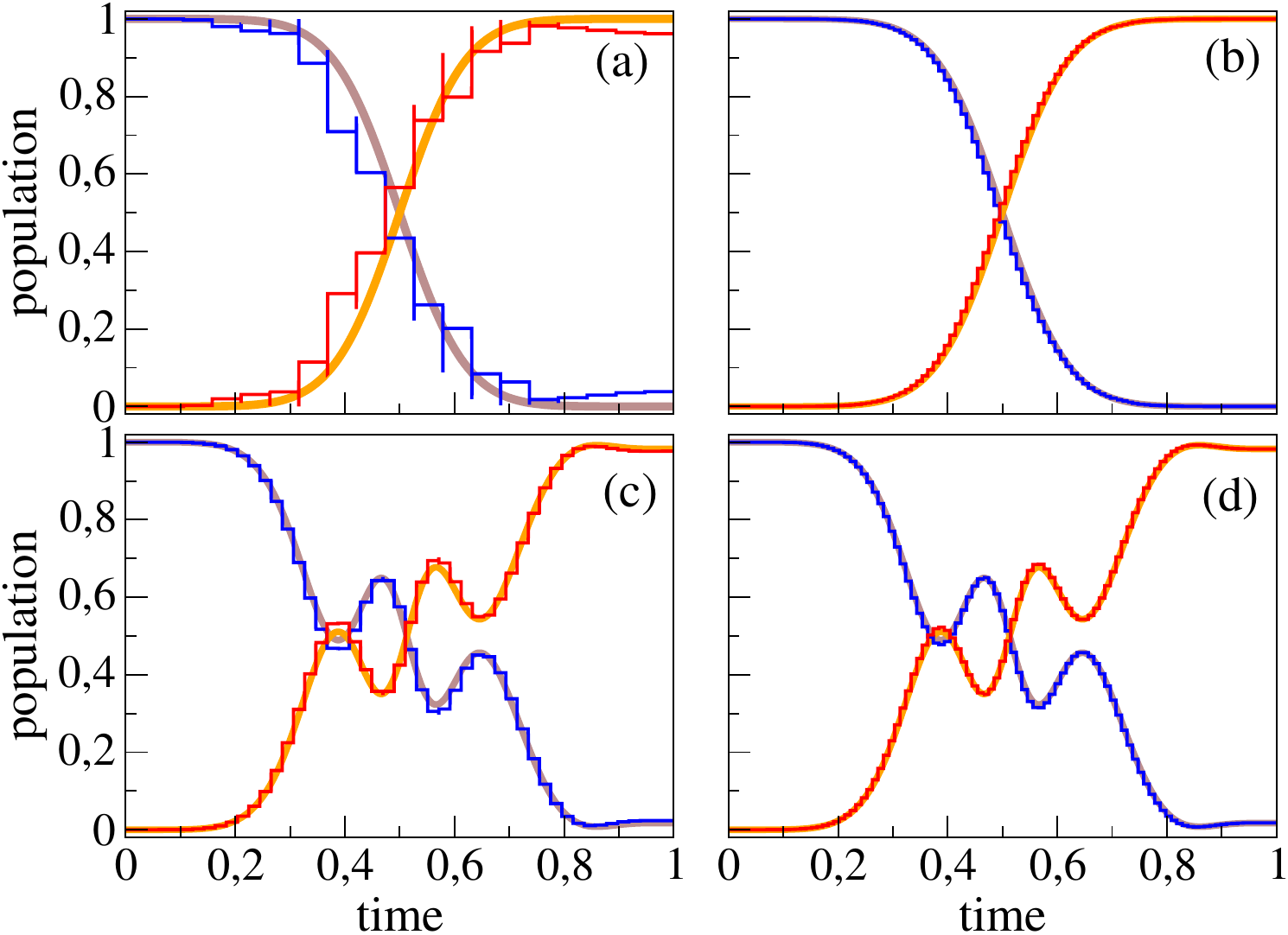}
\caption{Population dynamics by a continuous pulse with total pulse area of $5\pi$ and chirp $\alpha\tilde{\tau}^2 = 291.6$ (a) and (b) or $\alpha\tilde{\tau}^2 =64.8$ (c) and (d). The sampling (number of subpulses) in the digitized version is $N = 20, 100, 50$ and $100$ for subfigures (a) to (d) respectively.}
\label{fig:pop}
\end{figure}

As a reference, we chose $\tilde{A} = \tilde\Omega \tilde{\tau} S_0 = 5\pi$. The envelope function is of Blackman's shape (the Fourier synthesis of a Gaussian  function using the first two modes):
$S(t) = \Omega_0 \left[ 0.42 - 0.5\cos\left( 2\pi t{\tau}/\right) +
0.08 \cos\left( 4\pi t/{\tau}\right) \right]$ ($S_0 =0.42$).
We use chirped pulses with $\tilde{\tau} \tilde{\delta}(t) = \alpha \left(t- \tau/2\right)$, where $\alpha$ is the chirp rate, and the detuning is zero at the peak Rabi frequency. In Figs.~\ref{fig:pop}(a) and (b), $\alpha\tilde{\tau}^2 = 291.6$, leading to full and smooth adiabatic population transfer from $\tilde{P}_1(t)$ to $\tilde{P}_2(t)$, whereas in  Figs.~\ref{fig:pop}(c) and (d), $\alpha\tilde{\tau}^2 =64.8$, implying slow adiabatic conditions that lead to population oscillations.

\begin{figure}
\includegraphics[width=0.9\linewidth]{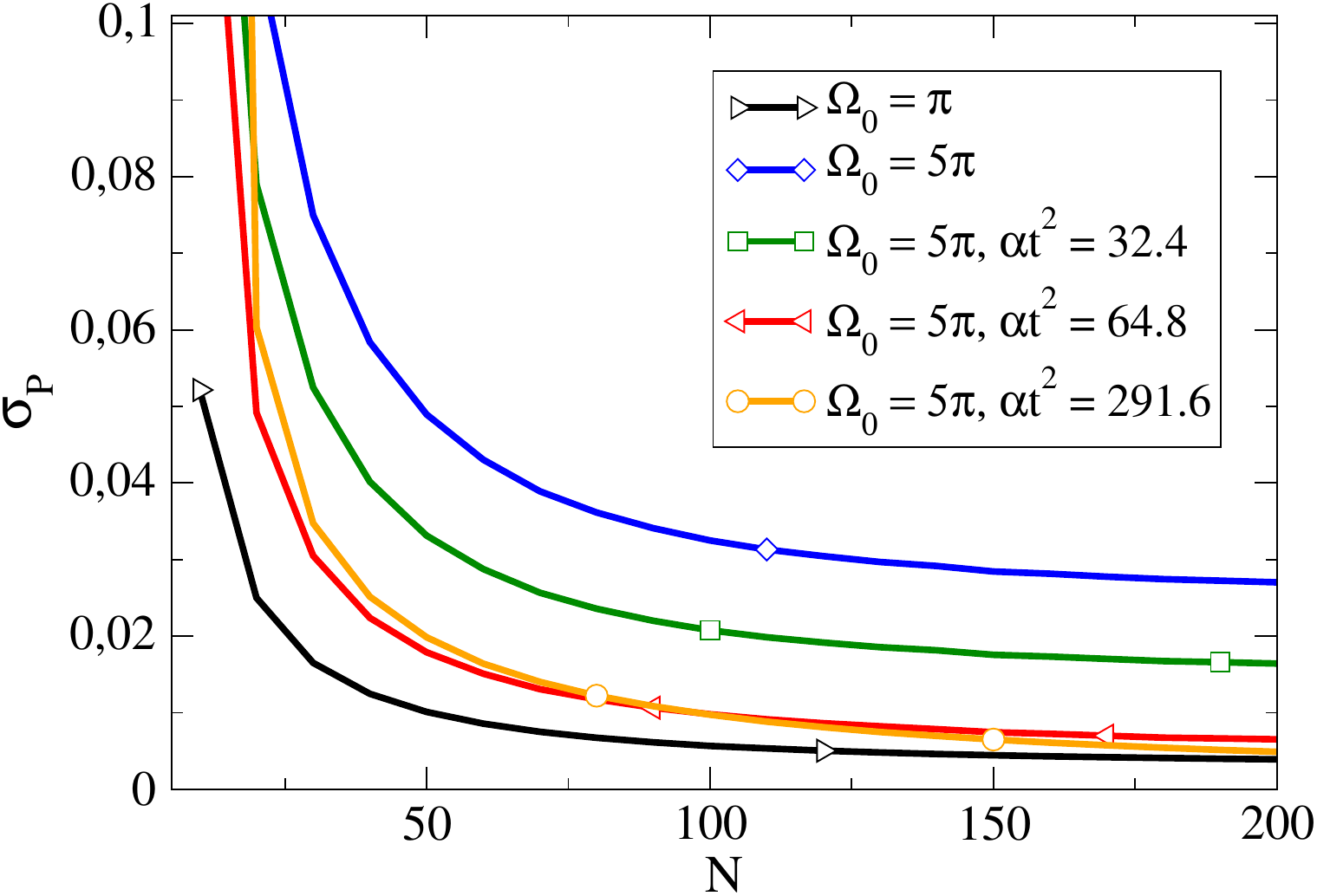}
\caption{Time-averaged population error, defined as square quadratic deviation of the population in the target state in the continuous $\tilde{P}_1(t)$ pulse from its digitized version $P_1(t)$, with $N = 100$ subpulses, for dynamics with different pulse areas and chirp ratios. The precision of the copy improves as the dynamics is more adiabatic, as it requires less subpulses to respond to all the population wiggles.}
\label{fig:Popdiff}
\end{figure}

For the dynamics under the pulse train, results are shown for different numbers of subpulses. In Fig.~\ref{fig:pop}(a), $N = 20$, in (b) $N = 100$, in (c) $N = 50$ and in (d) $N=100$. In general, we fix the ratio of the train period to the subpulse duration as $r_1 = 100$ except in Fig.~\ref{fig:pop}(c), where $r_1=50$. The populations show piecewise behavior that closely tracks the continuous dynamics (thin solid lines) for the long pulses, already with $N = 50$. For larger Rabi frequencies or dynamics with stronger oscillations, one needs a greater number of subpulses to follow correctly the oscillations. On the other hand, if the chirp is too large or the number of subpulses too small (such that the area of each subpulse must be large for the cumulative area of the train to be equal to the area of the ``continuous'' pulse), one can violate the conditions for the first-order perturbation approximation, and the dynamics under the pulse train deviates from the continuous dynamics. This is shown more clearly in Fig.~\ref{fig:Popdiff}, where we plot the integrated population differences as a function of the number of subpulses in the train,
\begin{equation}
    \tilde{\tau}\sigma_P = \sqrt{ \int_0^{\tilde{\tau}} \! dt
    \big( \tilde{P}_0(t) - P_0(t) \big)^2 } \ ,
\end{equation}
where $\tilde{P}_0(t)$, $P_0(t)$ are the populations of the initial state driven by the continuous pulse and by the pulse train (similar results would be obtained using the target state populations, $P_1(t)$). We have fixed $r_1 = 100$ in all cases. Deviations are smaller for the dynamics that show less oscillations ($\tilde{A} = \pi$, $\alpha \tilde{\tau}^2 = 291.6$), while those cases that exhibit higher oscillations need a larger number of subpulses to efficiently reproduce the dynamics ($\tilde{A} = 5\pi$, $\alpha\tilde{\tau}^2 = 32.4$). However, if the chirp is too large, as when $\alpha\tilde{\tau}^2 = 291.6$, the first-order perturbation conditions breakdown for $N < 20$, leading to large deviations from the desired dynamics.

Typically, for dynamics with a single Rabi flopping, the accumulated errors are below $1$\% already for $N\sim 100$ reaching values of the order of $2\cdot10^{-3}$ for $500$ subpulses. For other cases, one may need $200$ or more subpulses to achieve such fidelity. The dependence on $r_1$ is of the same type as the dependence with $N$. The area of the subpulses increases inversely with $r_1$. From the point of view of the approximations involved in the perturbation, halving $r_1$ is equivalent to halving $N$, as the areas of the subpulses are doubled and the stepwise increase or decrease of population with each subpulse is also doubled. However, while lowering $N$ makes for a larger step in the time axis (the $x$ axis in Fig.~\ref{fig:pop}), lowering $r_1$ produces the effect in the population axis (the $y$ axis in Fig.~\ref{fig:pop}).

\section{Sideband effects}

In this section we show that one can drive the desired transition from the initial to the target state in the digitized dynamics by tuning the frequency of the pulses in resonance with any sideband of the pulse train or frequency comb. This feature enhances the robustness of the dynamics with respect to detuning, and can be used for ``parallel'' excitation (using different sidebands) of different target excited states, creating controlled superposition states. Although the effect can be observed using trains with both frequency-varying and frequency fixed subpulses, here we concentrate on the latter for simplicity.

As the sidebands of the train are separated by $\omega_{k+1}-\omega_k = 2\pi T^{-1}$, one can recover population transfer at large detunings of the carrier frequency from the resonance, $\delta = | E_1 - E_0 -\omega | \ll \omega$ ($E_j$ are the energy levels and atomic units are used).
However, the spectra is not uniform: the intensity of the sidebands
decays depending on the Fourier spectra of the train's envelope function.
For full population transfer, one needs to increase the peak amplitude
of the pulse such that the pulse area at the sideband is an odd number
of $\pi$. We call ${\cal F}(\Omega_n)$ the amplitude of the train at the frequency corresponding to its $n$ sideband ($n$ is therefore the sideband order). We will focus on the dynamics with constant frequency trains, $\delta_{k+1} = \delta_k = \delta$.

\begin{figure}
\includegraphics[width=0.9\linewidth]{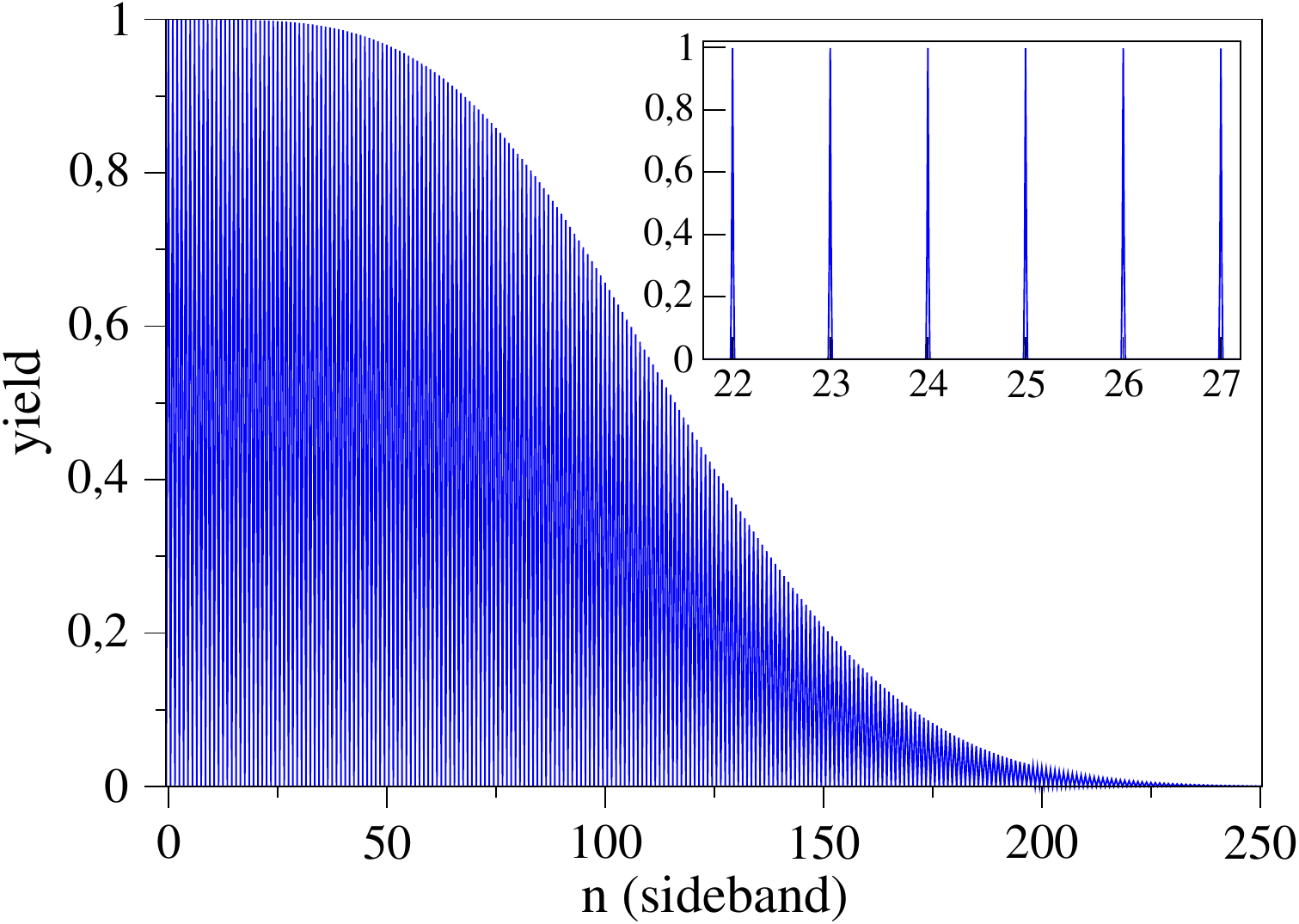}
\caption{Yield of population inversion at final time as a function of the energy difference $\Delta E$ between the levels, in units of the sideband of the train $n$. Whenever $T\Delta_E/2\pi$ is an integer number, there is resonance between a sideband of the train (for $n \neq 1$) and the molecular transition, leading to full population transfer. The yield decays for large $n$ due to the decay in the amplitude of the spectra with the sideband. The main figure shows the envelope of the yield. With higher resolution, the inset zooms in the yield when the transition is resonant with a few sidebands.}
\label{fig:sidebands}
\end{figure}

In Fig.~\ref{fig:sidebands} the yield of population transfer is measured as  final population in the excited state $|1\rangle$ at final time, $P_1(\infty)$. To facilitate  the comparison of the yield as a function of the detuning, we write the final population as $P_1(n)$, where in this
section the detuning $\delta$ is measured in units of the sideband order $n$ (that is, units of frequency are multiplied by $T/2\pi$).  Here, $n=0$ corresponds to the carrier frequency. The inset shows in detail the procession of full population transfer carried by the sidebands, following the sidebands spectra. In the results, we fix the peak Rabi frequency so that the pulse area at $n=0$ is $\pi$, and we use $N = r_1 = 100$. Hence,  
\begin{align}
P_1(n) = \sin\left( \frac{\pi}{2} \frac{{\cal F}(\Omega_n)}{{\cal F}(\Omega_0)}\right)^2 \ .
\end{align}
For Gaussian pulses, 
\begin{align}
P_1(n) = \sin\left( \frac{\pi}{2} e^{-\tau^2 n^2 / T^2}\right)^2 \ .
\end{align}
For Blackman shape pulses (and a large family of well-behaved pulse shapes) $P_1(n)$ is approximately $0.5$ at $n = N$.

\begin{figure}
\includegraphics[width=0.9\linewidth]{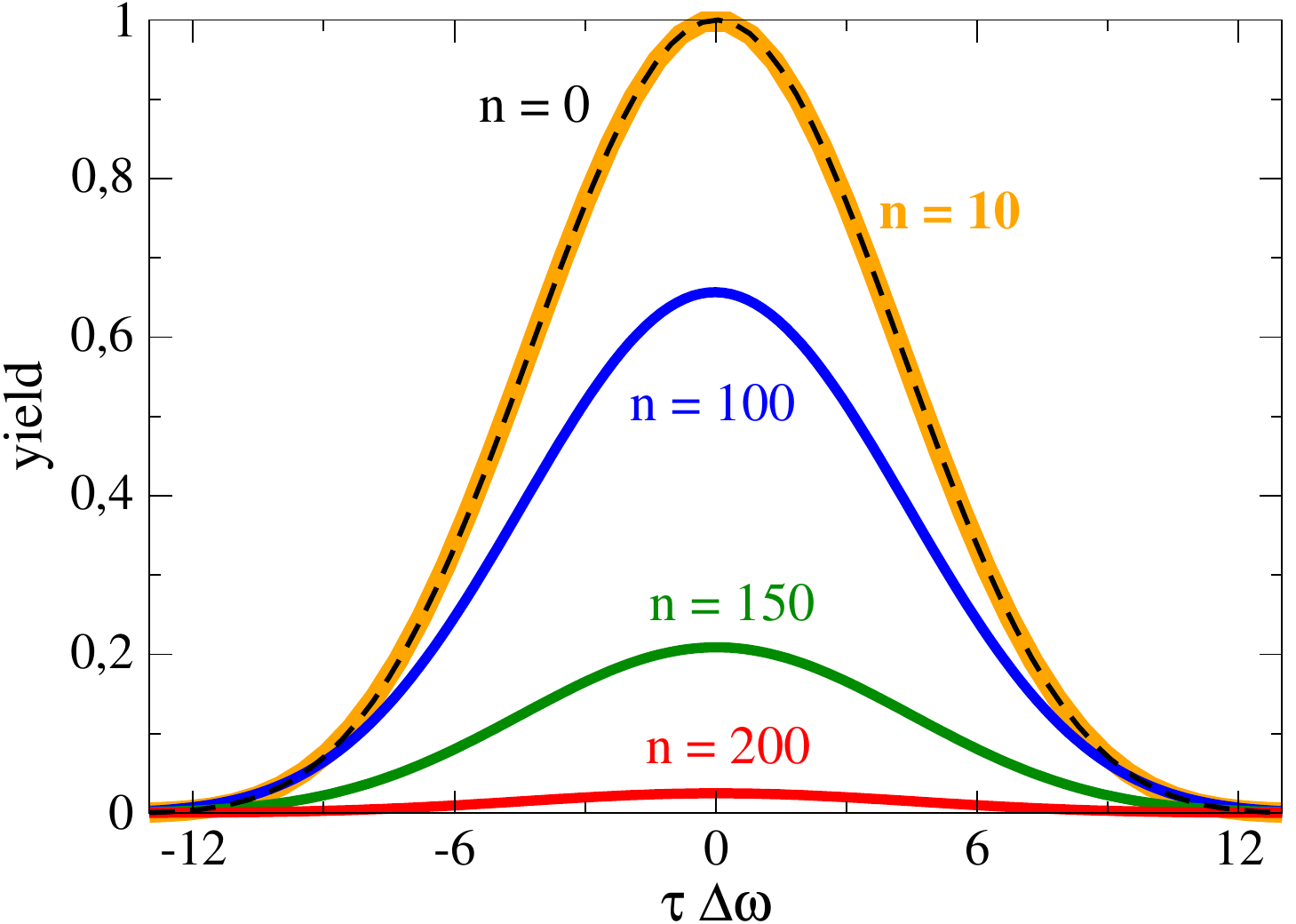}
\caption{Decay of the yield of population transfer as a function of the detuning from the resonant frequency for different sidebands, $n = 0, 10, 100, 150$ and $200$, which further zooms in the spikes shown in the inset of Fig.~4. Final populations follow the same behavior as a function of detuning, regardless of the sideband order.}
\label{fig:widths}
\end{figure}

Fig.~\ref{fig:widths} shows in detail how $P_1(n)$ decays around the
frequencies corresponding to different sidebands. The result for $n=0$ coincides with the result of the continuous dynamics. The decay profiles do not depend on the sideband order, as expected. The fact that one can excite the population of the excited state with different sidebands and the fact that the behavior as a function of $\delta$ around each sideband is exactly the same, can be used to prepare specific superposition states in a robust way, whenever two or more excited states are in resonance, at the same time, with two or more sidebands of the train.

\begin{figure}
\includegraphics[width=0.9\linewidth]{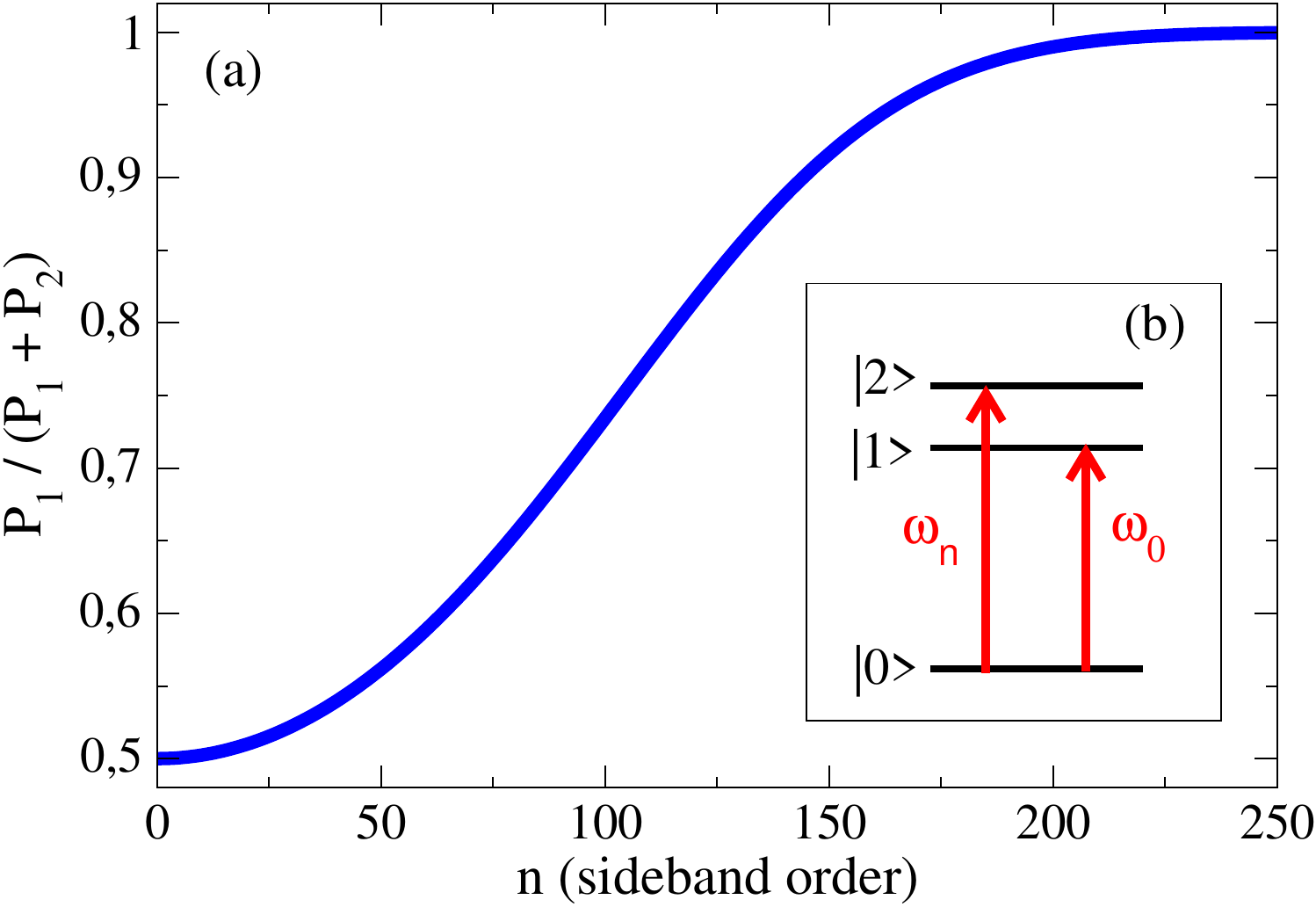}
\caption{Control of the relative populations in a superposition state. When two sidebands of the train, $0$ and $n$, are in resonance with two excited states $|1\rangle$ and $|2\rangle$ respectively, a superposition of the excited states is prepared with relative populations that depend on the relation between the order of the sidebands. In the inset (b) we show the diagram of the process.}
\label{fig:controlpop}
\end{figure}

We show this in Fig.~\ref{fig:controlpop}, where we add a second excited state $|2\rangle$, whose energy $E_2 = \omega + 2\pi n /T$ ($n \in \mathbb{Z}$), is exactly resonant with the $n$ sideband of the train. The figure shows the ratio of the final population in the first excited state, to the sum of populations in the excited manifold, $P_1 / (P_1 + P_2)$, where $|1\rangle$ is resonant with $n=0$ and $|2\rangle$ is resonant with $n$. We use a pulse train of $100$ subpulses with $r_1=100$ and with Rabi frequencies chosen according to Eq.(\ref{matchomega}).
The yield ranges between $0.5$ and $1.0$, within a $0.1$\% relative error.
Exchanging the role of $|1\rangle$ and $|2\rangle$ (so $|2\rangle$ is
now in resonance with $n=0$) gives control to the remaining range of
the yield from $0$ to $0.5$. 

As long as the period of the train matches the energy splitting of the levels, any deviation in the frequencies affects the populations in $P_1(0)$ and $P_2(n)$ in the same manner, hence not their ratio. Notice also that, in order to minimize the population in state $|0\rangle$
at final time, the peak Rabi frequencies must be adjusted for different
systems, as $n$ changes. If $E_2$ is resonant with the first sidebands
($n\sim 1-10$), the amplitude of the Rabi frequencies at the sidebands
is almost the same as for $n=0$ (see Fig.~\ref{fig:sidebands} or Fig.~\ref{fig:widths}). Because two states, and not one, are excited, the Rabi frequencies acting in the system are actually $\sqrt{2}\Omega_k$. Hence, for full population transfer, they should be divided by a $\sqrt{2}$ factor. On the other hand, if $E_2$ is resonant with the last sidebands $n \ge 200$, the amplitude at the sideband is much smaller than at $n=0$ and the Rabi frequency does not need to be adjusted. If two states are excited by sidebands $n$ and $m$, the actual Rabi frequency acting in the Hamiltonian is $\Omega'_k(t) = f_{n,m}  \Omega_k(t)$ with 
\begin{align}
f_{n,m} = \frac{\sqrt{{\cal F}(\Omega_n)^2+{\cal F}(\Omega_m)^2}}{{\cal F}(\Omega_0)}\,.
\end{align}
This pre-factor should be taken into account to achieve full population transfer. However, it does not affect the ratio of populations and therefore does not change the results of Fig.~\ref{fig:controlpop}.

\section{Conclusions}

We have demonstrated that a digitized version of Rapid Adiabatic Passage can be successfully implemented using a train of weak pulses with time-varying frequencies. By applying first-order perturbation theory to each subpulse excitation, we derived analytic conditions that allow the pulse train parameters (subpulse Rabi frequencies and detunings) to accurately reproduce the continuous-time dynamics of a traditional long-pulse excitation. The precision of the digitized version depends on the complexity of the population dynamics ({\it i.e.,} the population oscillations or wiggles), which is correlated with the adiabaticity of the process: dynamics similar to adiabatic passage require fewer subpulses in the train to accurately match the population history. On the other hand, if too few subpulses are used, the area of each pulse becomes too large, leading to a violation of the perturbative regime.

The pulse train serves as a weak-field analogue of strong-field adiabatic dynamics, providing comparable robustness to parameter variations while mitigating adverse effects such as optical damage or unwanted multiphoton processes. This analogy holds as long as the system does not enter a regime of very strong fields, where significant Stark shifts fundamentally modify the Hamiltonian. For two-level systems in which the population of the target state accumulates over the duration of the pulse train, the method is applicable only if the decay time from the target state significantly exceeds the train duration. In cases where this condition is not met, it may be possible to introduce corrections to the subpulses to compensate for population losses.

Furthermore, the inherent sidebands of the frequency comb introduce additional degrees of freedom, enabling resonant population transfer even when the carrier frequency is significantly detuned. By aligning specific sidebands with multiple excited states, the pulse train facilitates the preparation of precise superposition states. In this approach, control is limited to the populations rather than the phases. Therefore, for superpositions of two energy eigenstates, precise control of the overall process duration is necessary to determine the relative phase between the states.

In summary, the flexibility and coherence of pulse trains offer a versatile approach for controlling population dynamics, with significant potential for applications in emerging quantum technologies.

\section*{Acknowledgments}
This research was supported by Grant No. PID2021-122796NB-I00, 
funded by the MICIU/AEI/10.13039/501100011033 and the ERDF/EU, 
the MATRIX-CM project, TEC-2024/TEC-85, the DEVCOM Army Research 
Laboratory under Cooperative Agreement Number W911NF-24-2-0044
and by Grant No. NRF-2021R1A5A1030054, supported by the Center 
for Electron Transfer funded by the Korean government (MSIT). 
B.Y.C. acknowledges support from a María Zambrano grant
funded by the European Union - NextGenerationEU.

\bibliography{refs_new}

\end{document}